\documentclass[cameraready]{Interspeech}

\title{LMPAN: A Lightweight Multi-Path Alignment Network for Joint Full-Duplex Acoustic Echo Cancellation and Noise Suppression}

\author[affiliation={1},]{Chengwei}{Liu}
\author[affiliation={1,2}, correspondingauthor]{Shaofei}{Xue}
\author[affiliation={1},]{Haoyin}{Yan}
\author[affiliation={2}, ]{Xiaotao}{Liang}
\author[affiliation={2}, ]{Zheng}{Xue}

\address{
    $^1$ Qwen Business Unit of Alibaba, China \\
    $^2$ TongYi AI Lab of Alibaba Group, China 
}

\email{liuchengwei.lcw@alibaba-inc.com, mullerxue@126.com, yanhaoyin.yhy@alibaba-inc.com}


\usepackage{comment}

\usepackage{cite}
\usepackage{amsmath,amssymb,amsfonts}
\usepackage{algorithmic}
\usepackage{graphicx}
\usepackage{textcomp}
\usepackage{xcolor}

\usepackage{url}
\usepackage{color}
\usepackage{bm}
\usepackage{booktabs}
\usepackage{threeparttable}
\usepackage{verbatim}
\usepackage{multirow}
\usepackage{setspace}
\usepackage{etoolbox}

\usepackage{amssymb,amsfonts}
\usepackage{graphicx}
\usepackage{textcomp}
\usepackage{xcolor}
\usepackage{url}
\usepackage{color}
\usepackage{bm}
\usepackage{booktabs}
\usepackage{threeparttable}
\usepackage{verbatim}
\usepackage{multirow}
\usepackage{setspace}
\usepackage{etoolbox}
\usepackage{makecell}
\usepackage{amsmath,amssymb,amsfonts}
\usepackage{amsmath, amssymb, bm}
\usepackage{graphicx}
\usepackage{url}
\usepackage{hyperref}

\begin{document}

\maketitle

\begin{abstract}
We propose a lightweight multi-path alignment network (\textbf{LMPAN}) for on-device joint acoustic echo cancellation (AEC) and noise suppression (NS) in full-duplex spoken dialogue systems. To address hardware-induced distortions and dynamic acoustic conditions, we introduce three core innovations: (1) a multi-path alignment stage correcting temporal and energy mismatches across reference, linear AEC (LAEC) output, and microphone signals;
(2) an attention-based mechanism that dynamically integrates enhanced LAEC and microphone features under varying acoustic scenarios; (3) a post-filtering module with a dynamic target generation strategy for downstream tasks (ASR, VAD).
Furthermore, we adopt a two-stage training framework leveraging self-supervised learning representations to enhance perceptual quality. Experiments show that LMPAN, with only \textbf{480K} parameters and \textbf{126 MACs}, achieves performance comparable to the state-of-the-art lightweight model DeepVQE-S, while ensuring real-time inference capability.

\end{abstract}

\section{Introduction}
\label{sec:intro}
Full-duplex spoken dialogue systems (FDSDS) have made remarkable progress with the development of large language models (LLMs), enabling more natural interactions~\cite{LLM-FDSDS, LLMS-FDSDS}. 
However, their performance degrades substantially under adverse echo and noise conditions~\cite{sridhar2021icassp}, highlighting the critical importance of acoustic echo cancellation (AEC) and noise suppression (NS) for system efficacy~\cite{jiang2025icassp, benesty2001advances}. 

A variety of AEC approaches have been proposed, which can be broadly categorized into conventional digital signal processing (DSP)-based methods~\cite{benesty2001advances, cohen2021online} and neural network (NN)-based techniques~\cite{sridhar2021icassp,  jiang2024icassp,deevqe}. 
The performance of DSP-based methods is limited in complex full-duplex scenarios involving heterogeneous hardware, device-specific nonlinear distortions, time-varying latency (ranging from milliseconds to hundreds of milliseconds), and diverse acoustic environments~\cite{jiang2025icassp, heitkaemper2024improving, aec2023-challenge}.
In contrast, NN-based methods perform in-model alignment between the microphone (mic) and far-end signals, achieving superior accuracy and adaptability.

Within the NN-based approaches, end-to-end architectures have gained traction.
Evgenii et al.~\cite{deevqe} proposed DeepVQE for joint echo cancellation, noise reduction, and dereverberation (DRB). Yang et al.~\cite{FADI-AEC} introduced a diffusion-based stochastic regeneration approach, formulating AEC as a conditional generation problem. Other methods combine NNs with traditional signal processing components~\cite{two-step}, typically incorporating a linear AEC (LAEC) module based on adaptive filter algorithms to suppress linear echo components.
Additionally, multi-task learning frameworks~\cite{deevqe, Multi-Task, eskimez23_interspeech} have shown effectiveness in acoustic scenarios by jointly addressing NS, DRB, and AEC, leading to overall improvements in speech quality.
To address time-varying latency, prior works~\cite{deevqe, Align-ULCNet, NCA-CRN} proposed attention-based frameworks that integrate dynamic time alignment with parallel encoder structures to enhance AEC robustness. Some studies incorporate data augmentation and post-processing strategy to reduce the mismatch for downstream tasks such as automatic speech recognition (ASR) and voice activity detection (VAD)~\cite{jiang2025icassp, heitkaemper2024improving, braun2020data}.
\begin{figure}[t]
	\centering
	\includegraphics[scale=0.55]{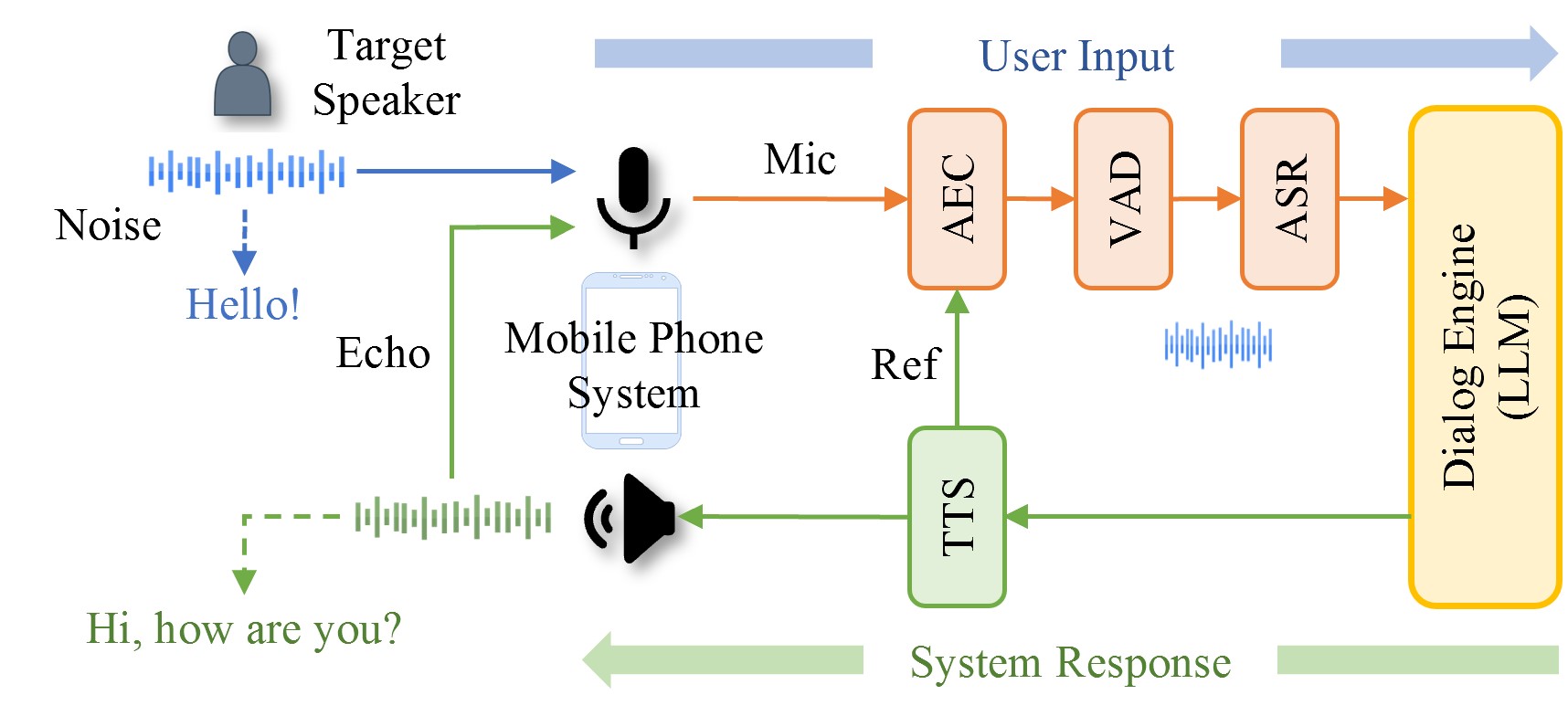}\vspace{-5pt}
	\caption{Full-Duplex Spoken Dialogue Architecture.}
	\label{fig:FDSDS_system}
	\vspace{-2em}
\end{figure}

Despite these advances, existing frameworks remain vulnerable to hardware-induced challenges in real-world FDSDS deployment. While capable of implicit alignment learning, these neural approaches lack explicit mechanisms to compensate for persistent temporal shifts and energy disparities across streams under varying hardware conditions. 
Consequently, such misalignments distort feature fusion, amplify residual artifacts, and ultimately degrade the fidelity required by sensitive downstream components~\cite{jiang2025icassp}. 
\begin{figure*}[!t]
	\centering
	\includegraphics[scale=0.49]{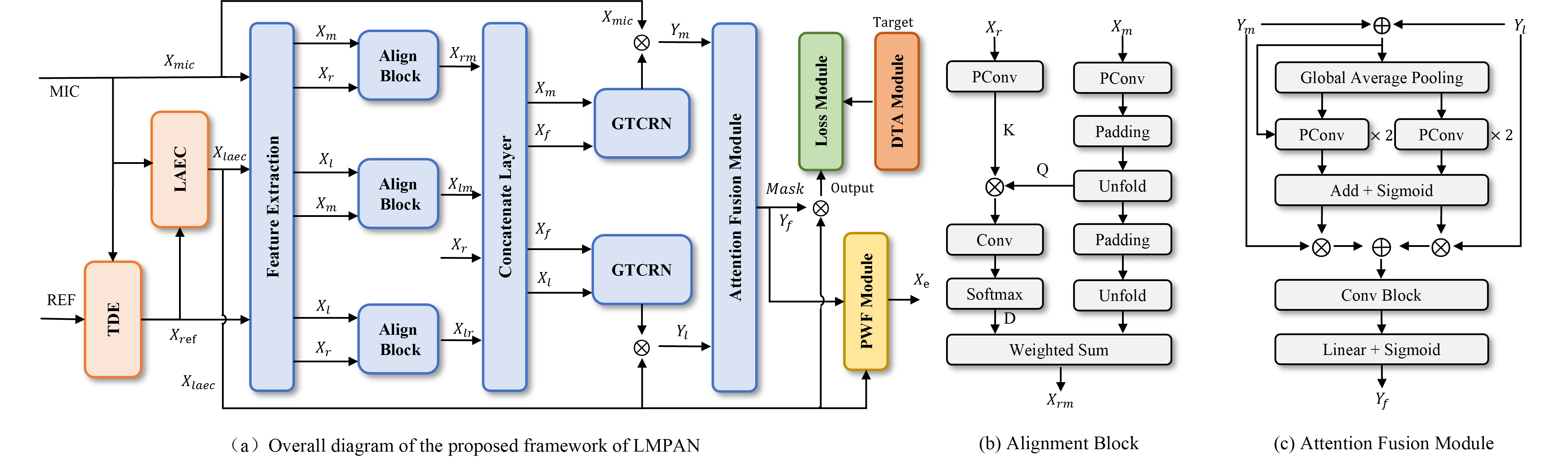}\vspace{-5pt}
	\caption{Overall structure of the proposed LMPAN system. Details for key components are given in: (a) Overall diagram of the proposed framework of LMPAN, (b) Alignment block, (c) Attention fusion module.}
	\label{fig:LMPAN_model}
	\vspace{-1.5em}
\end{figure*}

To address this gap, we propose LMPAN, a lightweight multi-path alignment network that holistically optimizes AEC performance through four synergistic innovations: (i) Dedicated multi-path alignment modules compensating temporal and energy mismatches; (ii) An attention-based adaptive fusion mechanism adapting to dynamic conditions without auxiliary VAD or double-talk detectors; (iii) A post-filtering module with dynamic target generation, which preserves speech integrity and prevents over-suppression to ensure robustness for ASR and VAD; and (iv) SSL-guided two-stage training leveraging WavLM representations for perceptual fidelity. 
\section{METHODOLOGY}
\label{sec:metho}


\subsection{Problem formulation}
We assume an FDSDS in Fig.~\ref{fig:FDSDS_system}, where near-end speech is contaminated by echo and noise. 
The observed signal model can be expressed as
\begin{equation}
	\label{eq1} \boldsymbol{y}=\boldsymbol{s}+\boldsymbol{n}+\boldsymbol{e}
\end{equation}
where $\boldsymbol{y}$, $\boldsymbol{s}$, $\boldsymbol{n}$, and $\boldsymbol{e}$ denote the microphone signal, near-end speech, additive noise, and echo component, respectively.
The echo component $\boldsymbol{e}$ is generated by convolving the far-end reference signal $\boldsymbol{r}$ with the acoustic echo path response $\boldsymbol{h}$. 
Given $\boldsymbol{r}$ as the reference signal, the target is to extract $\boldsymbol{s}$ from $\boldsymbol{y}$, involving both AEC and NS.

\subsection{Network Architecture}
Our proposed LMPAN is illustrated in Fig.~\ref{fig:LMPAN_model}. We compute the LAEC output by traditional algorithms, which is 
combined with mic and ref signals to serve as the input of neural network. 
To address temporal and energy misalignments among the LAEC output, mic signal, and ref signal, three pairwise alignment blocks are employed. These aligned features are then utilized to enhance the mic and LAEC signals, which are subsequently fused via an attention-based module. 
Finally, a post-processing strategy is applied to suppress nonlinear artifacts introduced by the neural network~\cite{jiang2025icassp}, with an optimal residual scaling parameter $\alpha = 0.4$ selected to minimize mismatch for downstream tasks (e.g., ASR, VAD). Module details follow. 

\noindent \textbf{LAEC:}
A sub-band time delay estimation (TDE) algorithm based on cross-correlation~\cite{azaria1984time} estimates the time delay between the ref and mic signals. A normalized least mean square (NLMS) based linear filter then estimates the linear echo component and outputs the residual signal.

\noindent \textbf{Multi-path Alignment:}
Conventional TDE methods degrade significantly under time-varying latency and low signal-to-noise ratio (SNR) conditions. Conversely, end-to-end neural approaches often lack explicit alignment mechanisms, leading to compromised robustness under data scarcity or domain shift. To address these limitations, the proposed alignment block (Fig.~\ref{fig:LMPAN_model}(b)) employs a soft temporal alignment strategy. 
Let $X_r \in \mathbb{R}^{2 \times t \times f}$ and $X_m \in \mathbb{R}^{2 \times t \times f}$ denote the power-compressed~\cite{li2021importance} complex 
short-time Fourier transform (STFT) spectrum
(real and imaginary parts are concatenated as 2-channel real-valued maps)
of far-end (ref) and mic signal, where $t$ and $f$ indicate time and frequency bins, respectively. 

First, feature maps are downsampled with a max-pooling layer (having a kernel size of $1 \times 4$) along the frequency dimension. Further, the features are reshaped such that $X_m'\in \mathbb{R}^{t \times (\frac{f}{4} \cdot c)}$ and $X_r'\in \mathbb{R}^{t \times (\frac{f}{4} \cdot c)}$. Queries $Q \in \mathbb{R}^{t \times p}$ and keys $K \in \mathbb{R}^{t \times p}$ are then projected from these representations, where $p \in \mathbb{N}$ is the projection size.
To estimate the time alignment between signals, a synthetic delay is applied to $K$ for each candidate delay $d \in [0, d_{\max}]$ by zero-padding at the start and cropping at the end. 
A dot product between $Q$ and the delayed $K$ yields a similarity score for each $d$, forming a delay likelihood vector of length $d_{\max}$. 
This vector is passed through a softmax to obtain a probabilistic delay distribution $D \in \mathbb{R}^{d_{\max}}$, where $d_{\max}=100$ corresponds to a maximum delay of 1 second, by zero-padding at the start and cropping at the end.
The aligned far-end features $\underline{X}_F \in \mathbb{R}^{2 \times t \times f}$ are computed via soft time alignment.

Three alignment blocks parallelly process ref and mic pair $[X_r, X_m]$, 
mic and LAEC pair $[X_m, X_l]$, and ref and LAEC pair $[X_r, X_l]$, yielding 
$X_{rm}$, $X_{ml}$, and $X_{rl} \in \mathbb{R}^{2 \times t \times f}$, respectively.
This soft alignment module accounts for uncertainty in temporal delays and enhances robustness to dynamic misalignments, eliminating the need for hard delay estimates.
Energy compensation employs learnable per-path scaling factors to normalize amplitude levels before fusion.

\noindent \textbf{Attention-based Fusion Module:}
Conventional NN-based AEC pipelines critically depend on LAEC outputs, yet LAEC inevitably introduces spectral distortions. To eliminate this vulnerability, we adopt an end-to-end dual-stream paradigm that jointly optimizes feature extraction from both the LAEC spectrum $Y_{l}$ and mic spectrum $Y_{m}$.

After alignment, $X_{rm}$, $X_{ml}$, $X_{rl}$, and $X_{r}$ are concatenated to $X_{f}$ along channel dimension, which is 
subsequently combined with $X_l$ and $X_m$ to refine LAEC spectrum and mic spectrum by GTCRN~\cite{rong2024gtcrn}, respectively. 
The attention fusion module Fig.~\ref{fig:LMPAN_model}(c) integrates information from refined LAEC spectrum $Y_{l}$ and mic spectrum $Y_{m}$, which captures
both \textit{local} and \textit{global} contextual features based on multi-scale channel attention mechanism.
The attention mask $M \in \mathbb{R}^{2 \times t \times f}$ and its complement $1 - M$ are applied to $Y_{l}$ and $Y_{m}$ as follows:
\begin{equation}
	Y_{f} = M \cdot Y_{l} + (1 - M) \cdot Y_{m}.
\end{equation}

\subsection{Dynamic Target Adaptation}
We propose a dynamic target generation strategy that adaptively retains controlled residual noise or echo to alleviate over-suppression in conventional AEC and NS systems.
To construct the training target $\boldsymbol{t}$, we introduce two residual scaling factors:
({\textit{a}}) A noise residual factor $\gamma$ controlled by a desired target signal-to-noise ratio (SNR):
\begin{equation}
	\gamma = \min\left(1,\ 10^{({\rm SNR}_{\rm in} - {\rm SNR}_{\rm t})/20}\right), 
\end{equation}
where ${\rm SNR}_{\rm in}$ and ${\rm SNR}_{\rm t}$ denote the input SNR and target SNR, respectively.
({\textit{b}}) An echo residual factor $\beta$ controlled by a desired target signal-to-echo ratio (SER):
\begin{equation}
	\beta  = \min\left(1,\ 10^{({\rm SER}_{\rm in} - {\rm SER}_{\rm t})/20}\right),
\end{equation}
where ${\rm SER}_{\rm in}$ and ${\rm SER}_{\rm t}$ indicate the input SER and target SER, respectively.
The final target signal is constructed as
\begin{equation}
	\boldsymbol{t} = \boldsymbol{s} + \gamma \boldsymbol{n'} + \beta \boldsymbol{e'},
\end{equation}
where $\boldsymbol{n'}$ and $\boldsymbol{e'}$ represent noise and echo components according to $\mathrm{SNR}_\mathrm{in}$ and $\mathrm{SER}_\mathrm{in}$, respectively. The factors $\gamma$ and $\beta$ retain a controlled level of interference, avoiding over-suppression and preserving realistic double-talk dynamics.
\begin{figure}[t]
	\centering
	\includegraphics[scale=0.40]{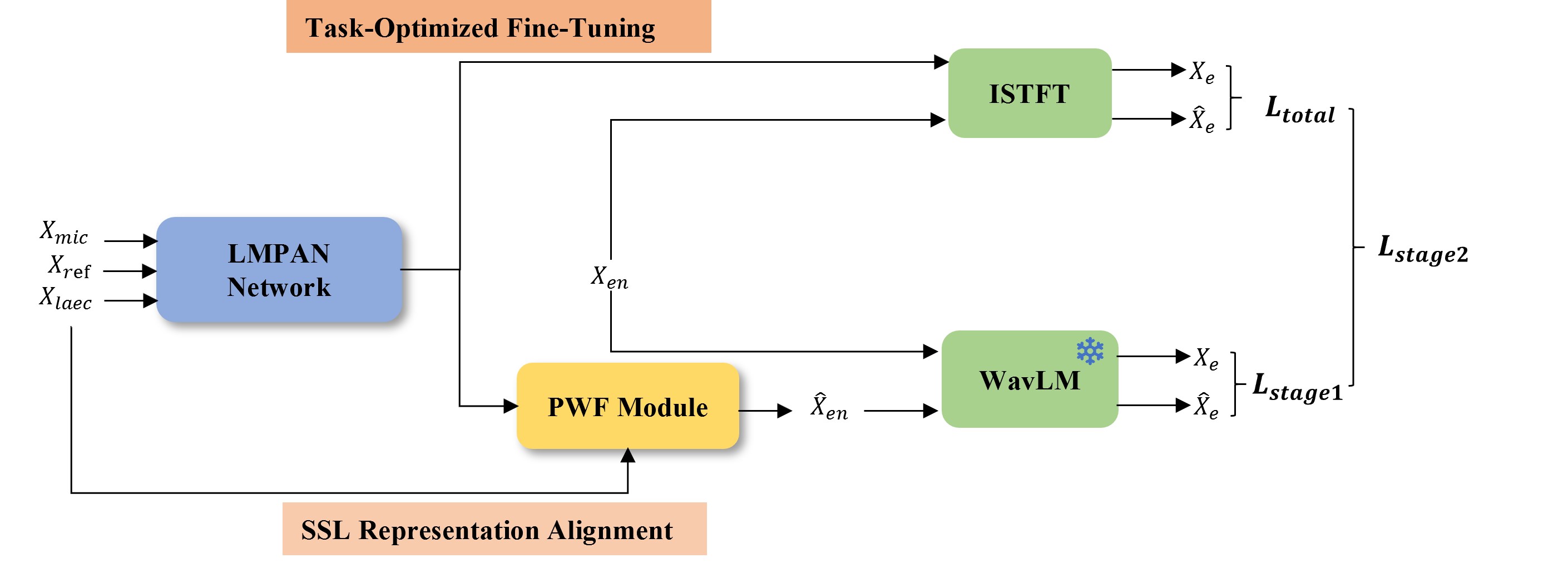}
	\caption{Two-stage training pipeline for LMPAN: Stage 1: SSL representation alignment using a frozen pretrained WavLM model; Stage 2 jointly optimizes spectral fidelity, echo suppression, and perceptual quality, with SSL loss as a consistency regularizer.}
	\label{fig:training}
	\vspace{-2.0em}
\end{figure}
\subsection{Loss Function}
Inspired by the effectiveness of self-supervised learning (SSL) embeddings in capturing rich acoustic and semantic representations~\cite{shankar2024closer, zhu2023vectok, li2025echofree}, we adopt a two-stage training strategy for LMPAN (Fig.~\ref{fig:training}) to jointly optimize representation alignment and task-specific objectives. The stage-wise objectives are formulated as:
\begin{align}
	\mathcal{L}_{\mathrm{stage\text{-}1}} &= \mathcal{L}_{\mathrm{SSL}}, \label{eq:stage1} \\
	\mathcal{L}_{\mathrm{stage\text{-}2}} &= 10 \, \mathcal{L}_{\mathrm{total}} + 0.5 \, \mathcal{L}_{\mathrm{SSL}} \label{eq:stage2},
\end{align}
\noindent\textbf{Stage 1: SSL Representation Alignment.} 
We freeze a pretrained WavLM-Large model~\cite{WavLM}\footnote{\url{https://huggingface.co/microsoft/wavlm-large}} and minimize the mean squared error (MSE) between SSL embeddings of the enhanced output and ground-truth clean speech: 
\begin{equation}
	\mathcal{L}_{\mathrm{SSL}} = \frac{1}{L} \sum_{l=1}^{L} \left\Vert \mathbf{e}_{l} - \hat{\mathbf{e}}_{l} \right\Vert^{2},
\end{equation}
where \(L\) denotes the total number of layers in the WavLM model.

\noindent\textbf{Stage 2: Task-Optimized Fine-Tuning.} We optimize a composite loss $\mathcal{L}_{\mathrm{total}}$ combining spectral reconstruction $\mathcal{L}_{\mathrm{spec}}$, echo-aware loss $\mathcal{L}_{\mathrm{echo}}$~\cite{Multi-Task}, scale-invariant SNR loss $\mathcal{L}_{\mathrm{si\text{-}snr}}$~\cite{SNR_LOSS}, and PMSQE perceptual loss $\mathcal{L}_{\mathrm{pmsqe}}$~\cite{PMSQE_LOSS}:
\begin{equation}
	\mathcal{L}_{\mathrm{total}} = \mathcal{L}_{\mathrm{spec}} + \alpha_1 \mathcal{L}_{\mathrm{echo}} + \alpha_2 \mathcal{L}_{\mathrm{si\text{-}snr}} + \alpha_3 \mathcal{L}_{\mathrm{pmsqe}},
	\label{eq:total_loss}
\end{equation}
where $\alpha_1 = 0.1$, $\alpha_2 = 0.2$, and $\alpha_3 = 0.8$ are weighting coefficients selected based on ablation studies. The SSL loss $\mathcal{L}_{\mathrm{SSL}}$ serves as a consistency regularizer during fine-tuning to preserve semantic fidelity. For ablation, we also evaluate an SSL-only variant (denoted as SSL) that uses only Stage~1 (\ref{eq:stage1}) without the second-stage fine-tuning, to isolate the contribution of two-stage training.

\begin{table*}[htbp]
	\centering
	\footnotesize
	\captionsetup{labelformat=empty}
	\caption{Table 1. AECMOS and ERLE (dB) results on AEC Challenge 2023 blind test sets. MA: multi-path alignment; AFM: attention-based fusion module; DTA: dynamic target adaptation; SSL-only: training with SSL loss only; STL: two-stage training with SSL.}
	\begin{tabular}{l l c c c c c c c c}
		\toprule
		\multirow{3}{*}{Exp} & \multirow{3}{*}{Method} & \multirow{3}{*}{\# Param.} & \multirow{3}{*}{MACs} & \multicolumn{6}{c}{AEC Challenge 2023} \\
		\cmidrule(lr){5-10}
		& & & & \multicolumn{3}{c}{DT} & \multicolumn{2}{c}{ST-FE} & \multirow{2}{*}{MOS$_{\text{avg}}$} \\
		\cmidrule(lr){5-7} \cmidrule(lr){8-9}
		& & & & EMOS & DMOS & ERLE & EMOS & DMOS & \\
		\midrule
		\multirow{3}{*}{--} & DeepVQE (E2E)~\cite{deevqe} & 0.82M & 315M & 4.62 & 4.02 & 65.7 & 4.61 & 4.36 & 4.40 \\
		& Align-ULCNet (Hybrid)~\cite{Align-ULCNet} & 0.69M & 100M & 4.60 & 3.80 & -- & 4.77 & 4.28 & 4.36 \\
		& TBNN (Hybrid)~\cite{two-step} & 9.56M & -- & 4.72 & 4.16 & -- & 4.70 & 3.91 & 4.37 \\
		\midrule
		1 & Base Model (One-stage) & 0.24M & 65M & 4.28 & 3.69 & 42.33 & 4.60 & 4.09 & 4.17 \\
		2 & +MA & 0.32M & 82M & 4.43 & 3.89 & 45.21 & 4.62 & 4.29 & 4.31 \\
		3 & +MA+AFM & 0.48M & 126M & 4.51 & 4.02 & \textbf{48.22} & 4.65 & 4.38 & 4.39 \\
		4 & +MA+AFM+SSL-only & 0.48M & 126M & 4.58 & 4.09 & 46.43 & 4.66 & 4.42 & 4.44 \\
		5 & +MA+AFM+STL & 0.48M & 126M & \textbf{4.63} & \textbf{4.17} & 47.15 & \textbf{4.71} & \textbf{4.44} & \textbf{4.49} \\
		6 & +MA+AFM+STL+DTA & 0.48M & 126M & 4.59 & 4.12 & 45.04 & 4.66 & 4.40 & 4.44 \\
		\bottomrule
	\end{tabular}
	\label{tab:aec_results}
	\vspace{-1em}
\end{table*}

\section{EXPERIMENTS}
\label{sec:exp}

\subsection{Experimental Setup}
\textbf{Datasets:} 
In our experiments, we utilize matched clean and noisy speech pairs from ICASSP 2022/2023 AEC Challenge~\cite{cutler2022icassp, aec2023-challenge}
and noise data from DNS Challenge~\cite{dubey2022icassp,reddy2020interspeech}.
For realistic full-duplex evaluation, we additionally collect a large-scale echo dataset from 40 smartphones
at varying playback volume levels (30\%--100\%), capturing diverse acoustic scenarios due to different hardware characteristics and software configurations.
Each device yields approximately 180 minutes of real-world interactions, including far-end reference and microphone signals.

\noindent \textbf{Simulation Details:}
Training data are augmented with room impulse responses (RIRs) simulated via the hybrid method~\cite{bezzam2020study}, reference signal perturbations including time-frequency masking and temporal shifts of 0--80~ms, dynamic utterance concatenation, and standard augmentations~\cite{jiang2024icassp, park2019specaugment}.
We generate 10,000 rooms (dimensions: $3\times3\times3$~m to $8\times5\times4$~m) and reverberation times (RT60: 0.2–1.2 s).
This yielded 2,000 hours of training data partitioned 8:1:1 across double-talk (DT), far-end single-talk (ST-FE), and near-end single-talk (ST-NE) scenarios.
SER in DT segments ranges from $-20$~dB to $15$~dB, and SNR in DT and ST-NE segments ranges from $-5$~dB to $25$~dB.
For evaluation, we adopt the AEC Challenge 2023 blind test~\cite{aec2023-challenge} and a self-collected real-world test set (~4,000 noisy DT utterances; loudspeaker volume: 60–100\%). All audio is resampled to 16k~Hz.

\begin{figure}[t]
	\centering
	\includegraphics[scale=0.49]{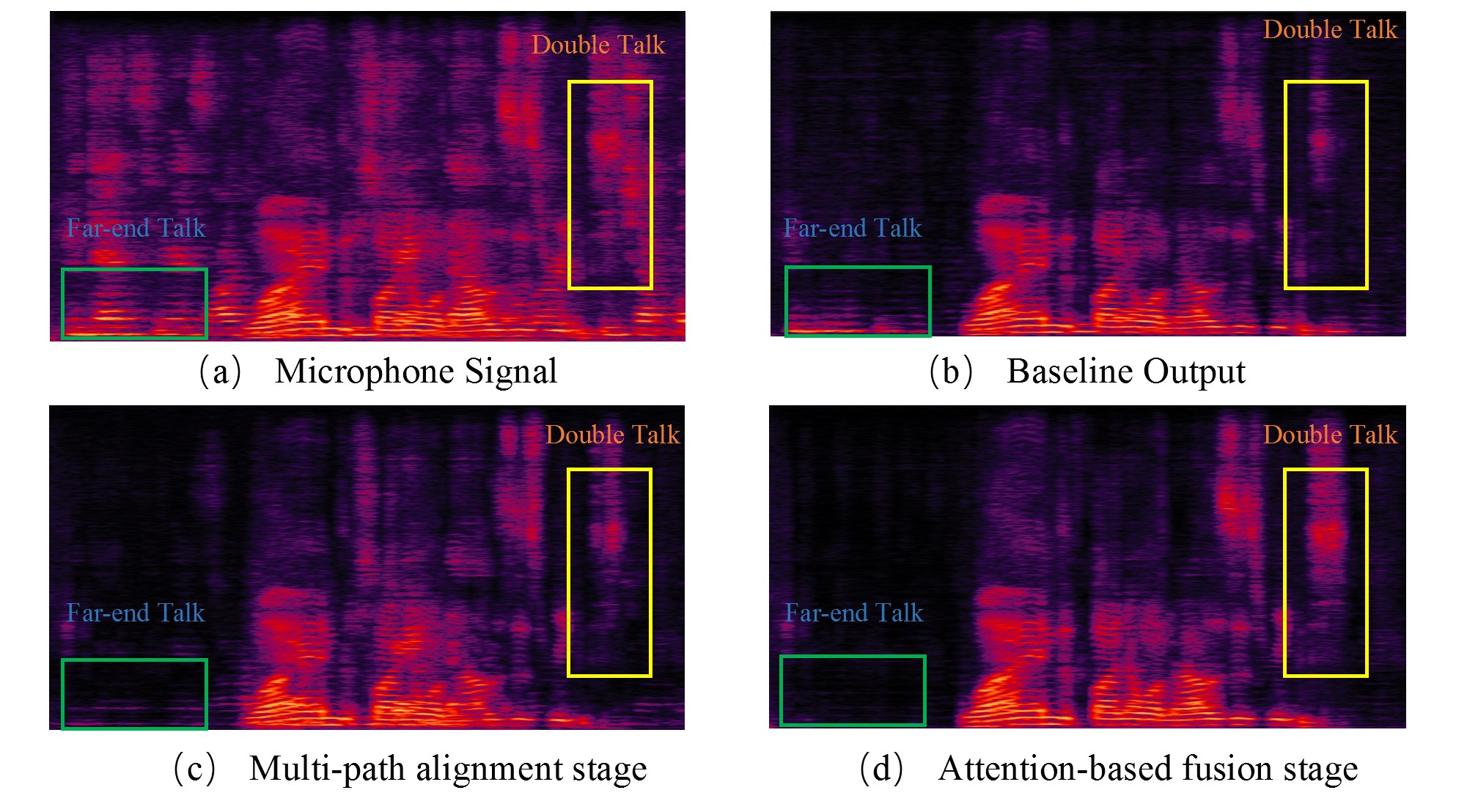}
	\vspace{-8pt}
	\caption{Spectrum visualizations of different stage on real double-talk sample of blind test.}
\label{fig:Vis_aec}
\vspace{-8pt}
\end{figure}

\noindent \textbf{Implementation Details:}
Input features were extracted via STFT with a 32-ms frame length, 16-ms frame shift, and magnitude compression factor of 0.3. 
This model is trained using the AdamW optimizer for 100 epochs. 
The learning rate reaches a peak of 0.001 after 4,000 warm-up steps and is decayed by a factor of 0.98 per epoch. 
The AFM employs $1\times1$ convolutions for Q/K/V projection with a single attention head. The GTCRN-based refinement branches use PConv with a $1\times3$ kernel along the frequency axis. During training, the ref and noisy signals are truncated to 5 seconds.

\noindent \textbf{Evaluation Metrics:}
For NS and AEC evaluation, the echo return loss enhancement (ERLE) and AECMOS~\cite{aecmos} are adopted in ST-NE and DT scenarios. AECMOS comprises two components: EMOS for echo annoyance and DMOS for other degradations, averaged as $MOS_{avg}$.
Downstream task evaluation is assessed via: (1) detection cost function (DCF) with semantic-VAD~\cite{shi2023semantic} model for VAD performance; (2) word error rate (WER) using Paraformer~\cite{gao2022paraformer}; (3) true interruption rate (TIR) via LLM-enhanced dialogue management~\cite{LLM-FDSDS} for FDSDS assessment.

\subsection{Experimental Results}

\noindent\textbf{AEC and NS Performance.}
Table~\ref{tab:aec_results} shows that LMPAN  achieves competitive performance on the AEC Challenge 2023 blind test. Starting from the lightweight base model (Exp\,1), integrating the multi-path alignment module (MA, Exp\,2) improves average MOS ($\text{MOS}_{\text{avg}}$) from 4.17 to 4.31. Incorporating the attention-based fusion module (AFM, Exp~3) increases $\text{MOS}_{\text{avg}}$ to 4.39 and ERLE to \textbf{48.22\,dB}. Two-stage training with SSL (STL, Exp\,5) further boosts perceptual quality to $\text{MOS}_{\text{avg}} = \mathbf{4.49}$, achieving the highest DT EMOS (4.63), DMOS (4.17), and ST-NE scores. Finally, DTA (Exp\,6) maintains robustness in double-talk but slightly reduces overall MOS. The final model (Exp\,5, 0.48M parameters, 126M MACs) outperforms prior methods including DeepVQE in $\text{MOS}_{\text{avg}}$ with lower complexity, demonstrating that architectural design, not scale, drives gains in efficient AEC systems.

\begin{table}[t]
	\centering
	\caption{Performance comparison of VAD (DCF\,\%), ASR (WER\,\%), and FDSDS (TIR\,\%) on a real-world double-talk test set, stratified by SER into $[-15, -10]$\,dB and $[-20, -15]$\,dB, with SNR randomly sampled from $[5, 20]$\,dB.}
	\label{tab:vad_asr}
	\resizebox{\columnwidth}{!}{%
		\begin{tabular}{l ccc ccc}
			\toprule
			\multirow{2}{*}{Method}
			& \multicolumn{3}{c}{SER: $[-15, -10]$\,dB}
			& \multicolumn{3}{c}{SER: $[-20, -15]$\,dB} \\
			\cmidrule(lr){2-4} \cmidrule(lr){5-7}
			& DCF\,$\downarrow$ & WER\,$\downarrow$ & TIR\,$\uparrow$
			& DCF\,$\downarrow$ & WER\,$\downarrow$ & TIR\,$\uparrow$ \\
			\midrule
			One-stage           & 4.68 & 11.08 & 90.96 & 9.38 & 24.25 & 85.17 \\
			\quad+\,MA          & 3.42 &  9.57 & 91.53 & 7.22 & 21.57 & 87.96 \\
			\quad+\,MA+AFM      & 2.55 &  8.36 & 92.82 & 5.85 & 19.38 & 89.17 \\
			\quad+\,MA+AFM+SSL-only  & 2.35 &  8.16 & 93.12 & 5.45 & 18.58 & 90.17 \\
			\quad+\,MA+AFM+STL  & 1.98 &  6.56 & 94.34 & 4.68 & 17.18 & 91.47 \\
			\quad+\,MA+AFM+STL+DTA  & \textbf{1.55} & \textbf{4.34} & \textbf{92.54} & \textbf{3.75} & \textbf{14.38} & \textbf{93.85} \\
			\bottomrule
		\end{tabular}%
	}
	\vspace{-1em}
\end{table}

\noindent\textbf{Downstream Task Performance.}
Table~\ref{tab:vad_asr} summarizes performance across VAD, ASR, and FDSDS tasks.
Compared to the one-stage baseline, the full pipeline (+MA+AFM+STL+DTA) achieves substantial gains in the more challenging $[-20, -15]$\,dB SER range (stronger echo interference), with DCF, WER, and TIR improving by 5.63, 9.87, and 8.68 percentage points, respectively.
Despite retaining controlled residual components, DTA and STL jointly boost performance by minimizing speech distortion, demonstrating the importance of target signal preservation in full-duplex systems.

\noindent\textbf{Ablation Studies.} Figure~\ref{fig:Vis_aec} visualizes spectrogram evolution across processing stages, demonstrating that integrating local microphone cues with multi-path alignment is essential for preserving near-end speech fidelity under strong echo interference. This is quantitatively validated in Table~\ref{tab:vad_asr}, where adding AFM (+\!MA $\rightarrow$ +\!MA+AFM) consistently improves ASR performance across all SER ranges, confirming that the LAEC+MIC fusion mechanism effectively preserves speech fidelity.

\noindent\textbf{Dynamic Target Adaptation Analysis.}
$\mathrm{SER}_\mathrm{t}$ is a hyperparameter defining residual echo retention in \textit{training targets} (not an output metric). Table~\ref{tab:beta_ablation} demonstrates that $\mathrm{SER}_{\mathrm{t}} = 25$\,dB optimizes WER for ASR by preserving moderate echo to minimize speech distortion. In contrast, $\mathrm{SER}_{\mathrm{t}} = 30$\,dB and $35$\,dB maximize PESQ and ERLE, respectively. This confirms that task-specific interference retention levels should be selected based on downstream objectives.

\begin{table}[t]
	\centering
	\caption{Ablation study of training target SER ($\mathrm{SER}_\mathrm{t}$) in dynamic target adaptation, evaluated on a simulated double-talk test set using WER\,(\%), PESQ, and ERLE\,(dB).}
	\label{tab:beta_ablation}
	\setlength{\tabcolsep}{10pt}
	\begin{tabular}{l ccc}
		\toprule
		$\mathrm{SER}_{\mathrm{t}}$\,(dB) & WER\,$\downarrow$ & PESQ\,$\uparrow$ & ERLE\,(dB)\,$\uparrow$ \\
		\midrule
		20 & 13.12          & 2.22          & 35.49 \\
		25 & \textbf{10.24} & 2.35          & 42.23 \\
		30 & 11.34          & \textbf{2.39} & 44.56 \\
		35 & 11.55          & 2.34          & \textbf{45.11} \\
		\bottomrule
	\end{tabular}
	\vspace{-1em}
\end{table}

\section{CONCLUSION}
In this paper, we propose LMPAN, a lightweight multi-path alignment network for on-device joint AEC and NS in full-duplex spoken dialogue systems.
By incorporating multi-path alignment and attention-based fusion module, 
the model effectively adapts to diverse acoustic conditions and hardware variations.
Combined with a two-stage training strategy leveraging SSL representations and a dynamic target adaptation mechanism that mitigates speech over-suppression, LMPAN preserves speech integrity critical for downstream tasks.
Experimental results demonstrate that LMPAN achieves competitive performance while maintaining low complexity and real-time inference, surpassing state-of-the-art lightweight baselines. Future work will focus on further optimizing the trade-off between resource efficiency and enhancement quality for end-to-end full-duplex spoken dialogue systems.

\bibliographystyle{IEEEtran}
\bibliography{refs}

\end{document}